\documentclass[conference]{IEEEtran}
\IEEEoverridecommandlockouts

\usepackage{acro}
\acsetup{single}

\DeclareAcronym{DeFi}{
  short = DeFi,
  long  = Decentralized Finance,
}
\newcommand{\DeFi}{\ac{DeFi}\xspace}

\DeclareAcronym{PoW}{
  short = PoW,
  long  = Proof-of-Work,
}
\newcommand{\PoW}{\ac{PoW}\xspace}

\DeclareAcronym{PoS}{
  short = PoS,
  long  = Proof-of-Stake,
}
\newcommand{\PoS}{\ac{PoS}\xspace}

\DeclareAcronym{APR}{
  short = APR,
  long  = Annual Percentage Rate,
}
\newcommand{\APR}{\ac{APR}\xspace}
\newcommand{\APRs}{\acp{APR}\xspace}

\DeclareAcronym{IL}{
  short = IL,
  long  = Impermanent Loss,
}

\DeclareAcronym{RL}{
  short = RL,
  long  = Realized Loss,
}

\DeclareAcronym{PNL}{
  short = PNL,
  long  = Profit and Loss,
}
\newcommand{\PNL}{\ac{PNL}\xspace} 
\newcommand{\PNLs}{\acp{PNL}\xspace}

\DeclareAcronym{LP}{
  short = LP,
  long  = Liquidity Provider,
}
\newcommand{\LP}{\ac{LP}\xspace}
\newcommand{\LPs}{\acp{LP}\xspace}

\DeclareAcronym{LT}{
  short = LT,
  long  = Liquidity Taker,
}

\newcommand{\LTs}{\acp{LT}\xspace}

\DeclareAcronym{NO}{
  short = NO,
  long  = Node Operator,
}

\DeclareAcronym{LSD}{
  short = LSD,
  long  = Liquid Staking Derivative,
}
\newcommand{\LSD}{\ac{LSD}\xspace}
\newcommand{\LSDs}{\acp{LSD}\xspace}

\DeclareAcronym{LST}{
  short = LST,
  long  = Liquid Staking Tokens,
}

\DeclareAcronym{DEX}{
  short = DEX,
  long  = Decentralized Exchange,
}
\newcommand{\DEX}{\ac{DEX}\xspace}
\newcommand{\DEXs}{\acp{DEX}\xspace}

\DeclareAcronym{CEX}{
  short = CEX,
  long  = Centralized Exchange,
}
\newcommand{\CEX}{\ac{CEX}\xspace}
\newcommand{\CEXes}{\acp{CEX}\xspace}

\DeclareAcronym{TVL}{
  short = TVL,
  long  = Total Value Locked,
}

\DeclareAcronym{DApp}{
  short = DApp,
  long  = Decentralized Application,
}

\DeclareAcronym{DAO}{
  short = DAO,
  long  = Decentralized Autonomous Organisation,
}

\DeclareAcronym{CeFi}{
  short = CeFi,
  long  = Centralized Finance,
}

\DeclareAcronym{MEV}{
  short = MEV,
  long  = Maximal Extractable Value,
}

\DeclareAcronym{SaaS}{
  short = SaaS,
  long  = Staking as a Service,
}
\newcommand{\SaaS}{\ac{SaaS}\xspace}

\newcommand{\USD}{\ensuremath{\xspace\texttt{USD}}\xspace}

\newcommand{\stETH}{\ensuremath{\xspace\texttt{stETH}}\xspace}

\newcommand{\ETH}{\texttt{ETH}\xspace}
\newcommand{\rETH}{\texttt{rETH}\xspace}
\newcommand{\frxETH}{\texttt{frxETH}\xspace}
\newcommand{\sfrxETH}{\texttt{sfrxETH}\xspace}
\newcommand{\sETHt}{\texttt{sETH2}\xspace}
\newcommand{\rETHt}{\texttt{rETH2}\xspace}

\newcommand{\tx}{\mathsf{tx}\xspace}

\usepackage[sort&compress, numbers]{natbib}
\usepackage{graphics}
\usepackage{hyperref}
\usepackage{amsmath}
\usepackage{xspace}
\usepackage{booktabs}
\usepackage{subfigure}
\usepackage{url}
\usepackage{mathtools}
\usepackage{amsfonts}
\usepackage{amssymb}
\usepackage{multirow}
\usepackage{enumitem}
\usepackage{numprint}
\usepackage{tcolorbox}
\usepackage{xurl}
\usepackage{xstring}
\usepackage[T1]{fontenc}
\usepackage{graphicx}
\usepackage{setspace}
\usepackage{makecell}
\usepackage{xcolor,colortbl}

\usepackage{caption}
\usepackage{algorithm}
\usepackage{algpseudocode}

\definecolor{gainsboro}{rgb}{0.86, 0.86, 0.86}
\definecolor{forest}{rgb}{0.86, 0.86, 0.86}

\newcommand{\xihan}[1]{{\color{black}{#1}}}

\definecolor{Gray}{gray}{0.9}

\usepackage{amssymb}
\usepackage{pifont}
\newcommand{\cmark}{\ding{51}}%
\newcommand{\xmark}{\ding{55}}%

\newcommand{\stETHPricePrimary}{$P_{\mathsf{stETH}}^{\mathsf{1st}}$\xspace}
\newcommand{\stETHPriceSecondary}{$P_{\mathsf{stETH}}^{\mathsf{2nd}}$\xspace}

\newcommand{\UST}{$\texttt{UST}$\xspace}
\newcommand{\LUNA}{$\texttt{LUNA}$\xspace}

\newcommand{\rETHPricePrimary}{$P_{\mathsf{rETH}}^{\mathsf{1st}}$\xspace}
\newcommand{\rETHPriceSecondary}{$P_{\mathsf{rETH}}^{\mathsf{2nd}}$\xspace}

\title{Exploring the Market Dynamics of \\ Liquid Staking Derivatives (LSDs)}

\author{
\IEEEauthorblockN{
Xihan Xiong\IEEEauthorrefmark{2}, 
Zhipeng Wang\IEEEauthorrefmark{2},
Qin Wang\IEEEauthorrefmark{3}
}

\IEEEauthorblockA{
\IEEEauthorrefmark{2}Imperial College London, United Kingdom\\
}
\IEEEauthorblockA{
\IEEEauthorrefmark{3}CSIRO Data61, Australia\\
}
}

\begin{document}

\maketitle

\begin{abstract}
Staking has emerged as a crucial concept following Ethereum's transition to Proof-of-Stake consensus. The introduction of Liquid Staking Derivatives (LSDs) has effectively addressed the illiquidity issue associated with solo staking, gaining significant market attention. This paper analyzes the LSD market dynamics from the perspectives of both liquidity takers (LTs) and liquidity providers (LPs). We first quantify the price discrepancy between the LSD primary and secondary markets. Then we investigate and empirically measure how LTs can leverage such discrepancy to exploit arbitrage opportunities, unveiling the potential barriers to LSD arbitrages. In addition, we evaluate the financial profit and losses experienced by LPs who supply LSDs for liquidity provision. Our results show that 66\% of LSD liquidity positions generate returns lower than those from simply holding the corresponding LSDs.

\end{abstract}

\section{Introduction}

Bitcoin uses the \PoW consensus (also referred to as Nakamoto consensus~\cite{bitcoin}) to achieve agreement among nodes in a decentralized setting. This consensus model was also adopted by smart contract-enabled blockchains such as Ethereum~\cite{wood2014ethereum}. While successfully maintaining network security, \PoW raised environmental concerns due to its substantial energy consumption. Consequently, the Ethereum community has been actively striving to propose more sustainable alternatives. Among these, \PoS~\cite{daian2019snow,buterin2020combining,gavzi2019proof,kiayias2017ouroboros} has risen as one of the most preferred options. Ethereum initiated its transition to a \PoS consensus on Dec~$1$,~$2020$, with the introduction of Beacon Chain. On Sep $15$,~$2022$, the Merge completes Ethereum's transition to \PoS consensus.

Subsequently, \PoS staking~\cite{zhang2023rationally} replaces \PoW mining on the Ethereum blockchain. Instead of relying on the computational power of \PoW mining to secure the network, \PoS depends on validators chosen to create new blocks based on the amount of \ETH they are willing to stake as collateral. Specifically, participants can lock up $32$ \ETH into the designated deposit contract to become validators. However, solo staking requires substantial capital commitment and technical expertise to maintain the validator node. Moreover, the staked \ETH becomes illiquid during the lock-up period, thus restricting users from capitalizing on broader market opportunities.

To address these challenges, the concept of \LSD emerged~\cite{scharnowski2023economics,xiong2023leverage,tzinas2023principal}. Liquid staking providers enable retail users, particularly those with limited capital and technical expertise, to collectively engage in the network's validation process and earn staking rewards. Various \LSDs adopt diverse token mechanisms to distribute staking rewards. For example, rebasing \LSDs adjust token supply to distribute staking rewards, whereas reward-bearing \LSDs increase token values to represent accumulated staking rewards. These token mechanisms not only determine reward distribution but also hold the potential to impact the dynamics of the \LSD market.

In the \LSD primary market, users can stake any desired amount of \ETH on liquid staking platforms to receive the corresponding \LSD, which represents both the underlying \ETH and the staking reward. Following this, users can utilize their \LSDs to integrate with existing \DeFi protocols in the secondary market. For instance, they can leverage \LSDs as collateral on lending platforms to borrow assets, contribute both \ETH and \LSDs to add liquidity to \DEX pools, and engage in asset trading by swapping \LSDs for other assets through \DEX pools.

Despite the considerable attention that LSDs have attracted, their market dynamics remain underexplored in existing literature. This paper aims to analyze \LSD market dynamics, focusing on the perspectives of both \LTs and \LPs. Our objectives are twofold. Firstly, we seek to investigate the price discrepancy between the \LSD primary and secondary market and understand how \LTs leverage this discrepancy for arbitrage opportunities. Secondly, we aim to evaluate the financial profit and losses experienced by \LPs who engage in supplying \LSDs for liquidity provision. We outline the main contribution of this paper as follows.

\begin{itemize}
    \item[$\diamond$]  \textbf{Token Mechanisms Systematization.} We systematically categorize the token mechanisms implemented by \LSD protocols, including the rebasing, reward-bearing, and dual-token models. We further formalize token mechanisms specific to rebasing \LSD (e.g., \stETH) and reward-bearing \LSD (e.g., \rETH), illustrating their functionality in the distribution of staking rewards.

    \item[$\diamond$] \textbf{Price Discrepancy and Arbitrage Analysis.} For rebasing and reward-bearing \LSDs, we quantify their price discrepancies between primary and secondary markets. Furthermore, we identify  $325.6$k~\ETH ($683.8$m~\USD) arbitrage amount caused by such price discrepancies since the inception of \LSDs. We provide empirical insights into the strategies adopted by arbitrageurs, revealing potential entry barriers in the context of arbitrages with \LSDs.


    \item[$\diamond$] \textbf{\LSDs Liquidity Provision Measurement.} We empirically measure the \APR experienced by $1{,}002$ \LPs who supplies \LSDs (e.g., \stETH and \rETH) to \DEX liquidity pools including Curve, Uniswap V3, and Balancer. We find that $66\%$ of \LSD liquidity provision positions yield a net \APR lower than the \APR of simply holding the corresponding \LSDs.
\end{itemize}

\section{Backgroud}

    \subsection{Blockchain and DeFi}

    Permissionless blockchains are decentralized distributed ledgers overlaying a global peer-to-peer network infrastructure, allowing any entity to join and participate freely. Within this context, especially in systems such as Ethereum~\cite{wood2014ethereum}, participants can create many decentralized applications using smart contracts. Built upon permissionless blockchains, \DeFi~\cite{werner2022sok, jiang2023decentralized} empowers users to participate in decentralized financial activities such as lending, borrowing, and trading.

    \subsection{From Pow Mining to PoS Staking} 
    \
    \PoW consensus has also been adopted by Ethereum since its inception. Nevertheless, a notable drawback of \PoW lies in its substantial computational demands and energy consumption. Motivated by the need for a more sustainable consensus mechanism, Ethereum embarked on a transition from \PoW~\cite{wood2014ethereum} to \PoS~\cite{grandjean2023ethereum,schwarz2022three,agrawal2022proofs,tang2023transaction, kapengut2023event}. This transition started in Dec~$2020$ by introducing the Beacon Chain system with a ``staking''  mechanism. Participants can deposit $32$ \ETH into the designated contract, thereby taking on the validator role for block proposal, block attestation, and synchronizing committee~\cite{grandjean2023ethereum}. On Sep $15$,~$2022$, the Ethereum ``Merge'' formally adopted the Beacon Chain as the new consensus layer to the original Mainnet execution layer. After the Merge, staking on the \PoS system replaces \PoW mining, reducing energy consumption by an estimated $99.95\%$~\cite{EthMerge2022}. On Apr $12$, $2023$, Ethereum underwent the ``Shapella upgrade'', facilitating the withdrawal of the staked \ETH for validators~\cite{EIP-4895}.

    \subsection{Staking Options on Ethereum}

    Ethereum participants are presented with four staking options: solo staking, \SaaS, pooled staking, and \CEX staking (see \ Fig.~\ref{fig:staking_options}).

    In \emph{solo staking}, participants manage their validator nodes by staking at least $32$ \ETH, ensuring complete control over staking rewards. This staking strategy can enhance blockchain security. However, it requires technical expertise in operating a validator node and a large capital commitment of a full $32$ \ETH, presenting a significant barrier to user participation.

    Compared with solo staking, \emph{\SaaS staking} substantially reduces the operational burden for users who possess $32$ \ETH but lack technical expertise.  The \SaaS provider manages the validator node on behalf of the user, receiving operational fees proportional to the amount of staked \ETH and staking rewards.

    Users with holdings below $32$ \ETH can choose \emph{pooled staking}, a system where multiple participants combine (or ``pool'') their \ETH to participate in the staking process collectively without necessitating individual full-node commitments. As such, all rewards and penalties accrued by the pool's validators are shared among stakers. Typically, staking pools charge fees as a fixed amount or a percentage of the staking rewards.

    In \emph{\CEX staking}, providers such as Binance and Coinbase offer users centralized and custodial staking services. Staking through \CEXes offers users the simplicity and convenience akin to pooled staking, eliminating the need for technical requirements and full-node commitments.

\section{Liquid Staking Derivatives}

Staking \ETH on Ethereum helps enhance network security and generates staking rewards. However, it restricts liquidity during the staking period, limiting users' ability to capitalize on market opportunities. In addressing this challenge, the concept of \LSD emerged, which serves as a tradable representation encompassing the underlying staked \ETH, its associated staking rewards and potential penalties (e.g., slashing~\cite{cassez2022formal}). Users can acquire \LSDs by participating in pooled staking (e.g., \href{https://lido.fi/}{Lido}) or \CEX staking (e.g., \href{https://www.coinbase.com/earn}{Coinbase}). These \LSDs can be traded instantly in the secondary market. As of the latest update, liquid staking protocols on Ethereum have accumulated a total value exceeding $20$b \USD\footnote{\url{https://defillama.com/protocols/liquid\%20staking/Ethereum}, access on Oct $5$, $2023$}, securing a leading position across various \DeFi sectors.

 \begin{figure}[t]
    \centering
    \includegraphics[width=0.88\columnwidth]{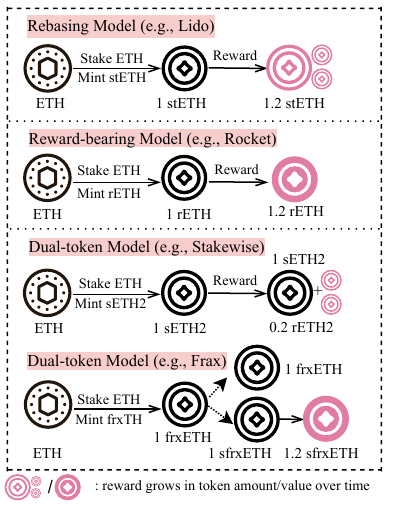}
    \caption{Illustration of LSD token mechanisms.}
    \label{fig: token_models}
 \end{figure}

\LSD token mechanisms can be broadly classified into the following categories: \emph{(i)} rebasing model, \emph{(ii)} reward-bearing model, and \emph{(iii)} dual-token model~(see Fig.~\ref{fig: token_models}).

\subsection{Rebasing Model} 

Tokens with a rebasing mechanism feature an elastic total supply that can increase or decrease, with the change in supply distributed proportionally among token holders. Stakers' \stETH balances get adjusted daily to reflect the accumulated rewards. The rebase can be positive or negative, depending on the validators' performance. The rebasing model mitigates the expense associated with reward distribution. Distributing staking rewards among all \stETH holders using direct transfer calls necessitates an unbounded loop. With rebasing, the \stETH smart contract can automatically update all the addresses holding \stETH in a single transaction. However, a rebasing token is difficult to integrate into existing \DeFi protocols, as the token's supply changes on a regular basis.

For instance, users who stake $m$ \ETH on \href{https://lido.fi/}{Lido} can obtain $m$ \stETH. Note that \stETH rebases via the ``share'' concept. Given the share price $p_{share}(t_0)$ at time $t_0$, a rebase event at time $t_1$ changes the share price to $p_{share}(t_1)$, consequently adjusting users' \stETH balances from $\frac{m}{p_{share}(t_0)}$ to $\frac{m}{p_{share}(t_1)}$. 
\begin{equation}\label{eq:rebasing}
   {\small
    \begin{split}
         &p_{share}(t_0) = \frac{\mathsf{totalETH}(t_0)}{\mathsf{totalShares}(t_0)} \\
        &p_{share}(t_1) = \frac{\mathsf{totalETHWithRewards}(t_1)}{\mathsf{totalShares}(t_0) + \mathsf{shares2mint}(t_1)}\\
        &\mathsf{shares2mint}(t_1) = \\
        &\frac{\mathsf{rewards} \cdot \mathsf{protocolFee} \cdot \mathsf{totalETH}(t_0)}{\mathsf{totalETHWithRewards}(t_1)- \mathsf{rewards} \cdot \mathsf{protocolFee}}
    \end{split}
    }
\end{equation}

\subsection{Reward-Bearing Model}

Reward-bearing \LSDs increase in their values to reflect the accumulated rewards. In contrast to rebasing \LSDs, reward-bearing \LSDs adopt an alternative design to simplify liquid staking while upholding stakers' seamless access to \DeFi opportunities, thus striking a balance between accessibility and functionality. Their supply remains stable, offering a more consistent valuation trajectory. For instance, \href{https://rocketpool.net/}{Rocket Pool} offers \rETH, which represents the tokenized staking assets and the rewards it gains over time. Notably, as staking rewards are earned, the value of \rETH appreciates, manifesting through changes in the \rETH/\ETH ratio at time $t$ (see Eq.~\ref{eq:reward_bearing}), while the holder's \rETH balance remains unchanged.

\begin{equation} \label{eq:reward_bearing}
    {\small
    \begin{split}
    \mathsf{P_{rETH/ETH}}(t) = \frac{\mathsf{totalETHStaked}_{t}+\mathsf{stakingRewardInETH}_{t}}{\mathsf{rETHTotalSupply}_{t}}
    \end{split}
    }
\end{equation}

\subsection{Dual-token Model}

The dual-token model entails two variations of \LSDs: \emph{(i)} a base token representing the underlying \ETH token on a 1:1 basis; and \emph{(ii)} a reward-bearing token that progressively accrues yield, or a reward token held separately by stakers to reflect the net reward. For instance, \href{https://app.frax.finance/staking/overview}{Frax} implements the dual-token model with its \frxETH and \sfrxETH tokens, where \frxETH maintains parity with \ETH and \sfrxETH accrues the staking reward. Stakers can choose between holding \frxETH to yield from liquidity provision in Curve's \frxETH--\ETH liquidity pool, or exchange \frxETH for \sfrxETH to earn the staking reward. In contrast, \href{https://stakewise.io/}{Stakewise} implements a different design, where the balance of \ETH deposits and rewards is reflected in \sETHt (staking \ETH) and \rETHt (reward \ETH) minted to stakers in a 1:1 ratio. This design avoids rebasing or reward-bearing dynamics, thereby mitigating the potential for impermanent loss when providing liquidity in \DEXs.



\section{LSD Price Discrepancy and Arbitrages}

    \begin{figure}[t]
    \centering
    \includegraphics[width=0.9\columnwidth]{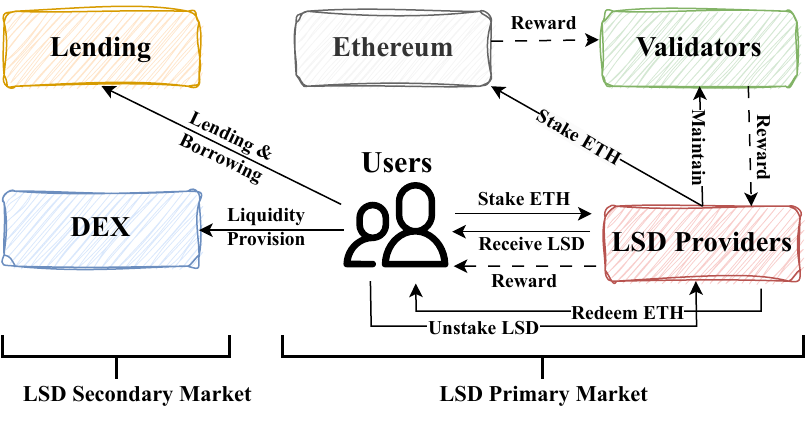}
    \caption{The primary and secondary markets for \LSDs.}
    \label{fig: lsd_market}
    \end{figure}

Users can stake \ETH on liquid staking platforms to acquire \LSDs in the primary market. These \LSDs can be utilized to integrate with existing \DeFi protocols in the secondary market, including participating in liquidity provision within \DEX pools and providing assets on lending platforms (see Fig.~\ref{fig: lsd_market}).

We first examine the price discrepancies between LSD primary and secondary markets, highlighting arbitrage opportunities. We then provide empirical insights into these behaviors and identify potential entry barriers for LSD arbitrage.

\subsection{LSD Price Discrepancy}

LSDs with varying token mechanisms show different price dynamics. In the subsequent sections, we do not discuss dual-token \LSDs due to their relatively low market share.

\begin{figure*}[t]
    \centering
    \includegraphics[width=0.88\linewidth]{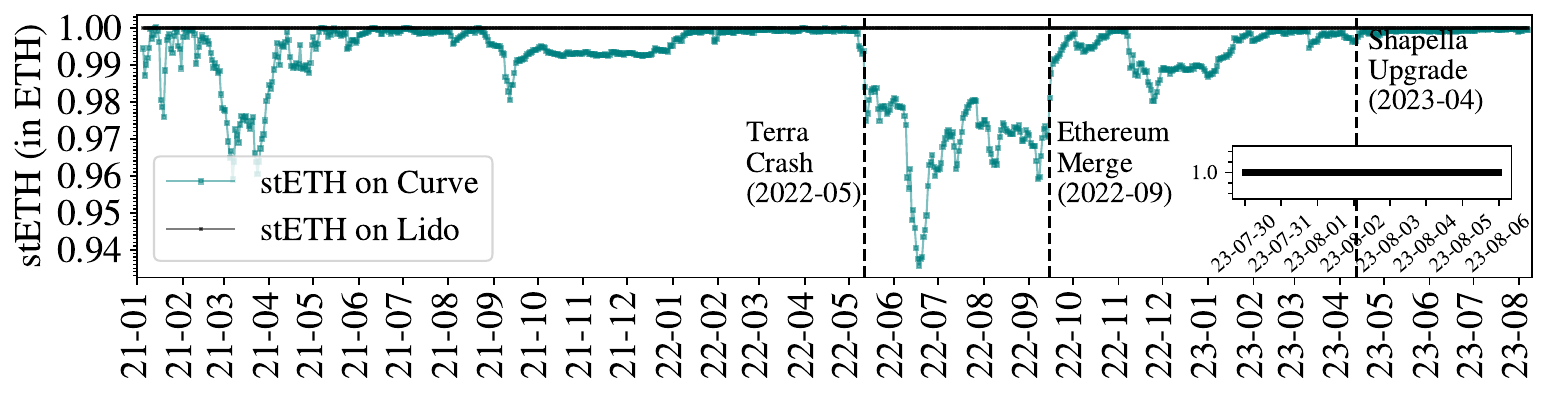}
    \caption{\stETH price on different platforms over time. This figure is adapted and extended from~\cite{xiong2023leverage}.}
    \label{fig: lido_wstETH_price_over_time}
\end{figure*}

\begin{figure*}[t]
\centering
\begin{minipage}{.49\textwidth}
  \centering
  \includegraphics[width=0.92\linewidth]{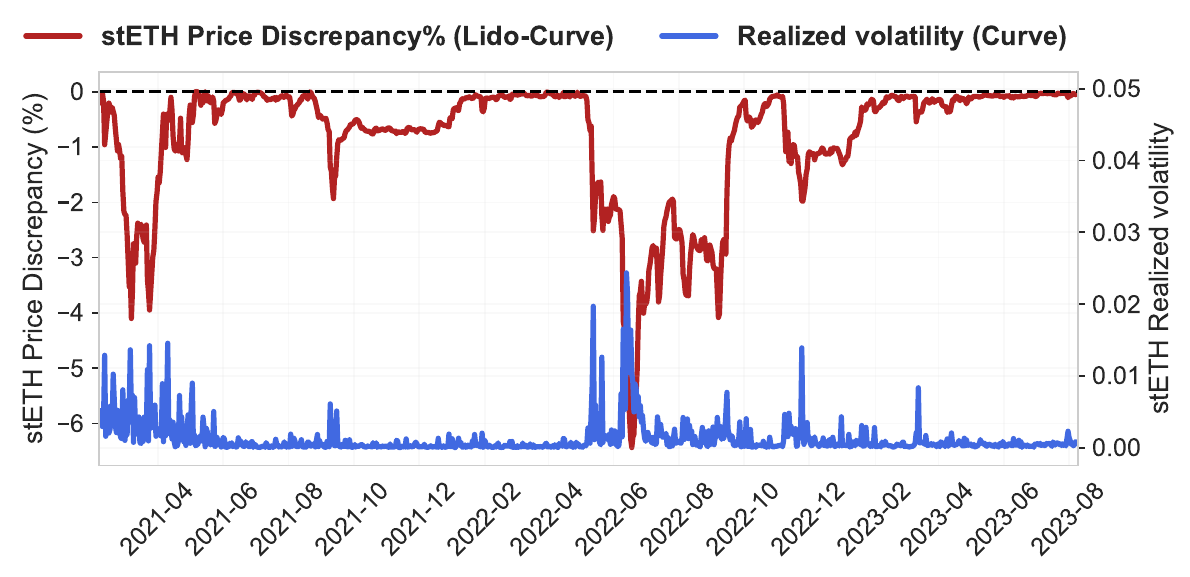}
    \caption{\stETH price discrepancy and realized volatility.}
    \label{fig:stETH_diff_vol}
\end{minipage}%
\begin{minipage}{.49\textwidth}
  \centering
    \includegraphics[width=0.9\linewidth]{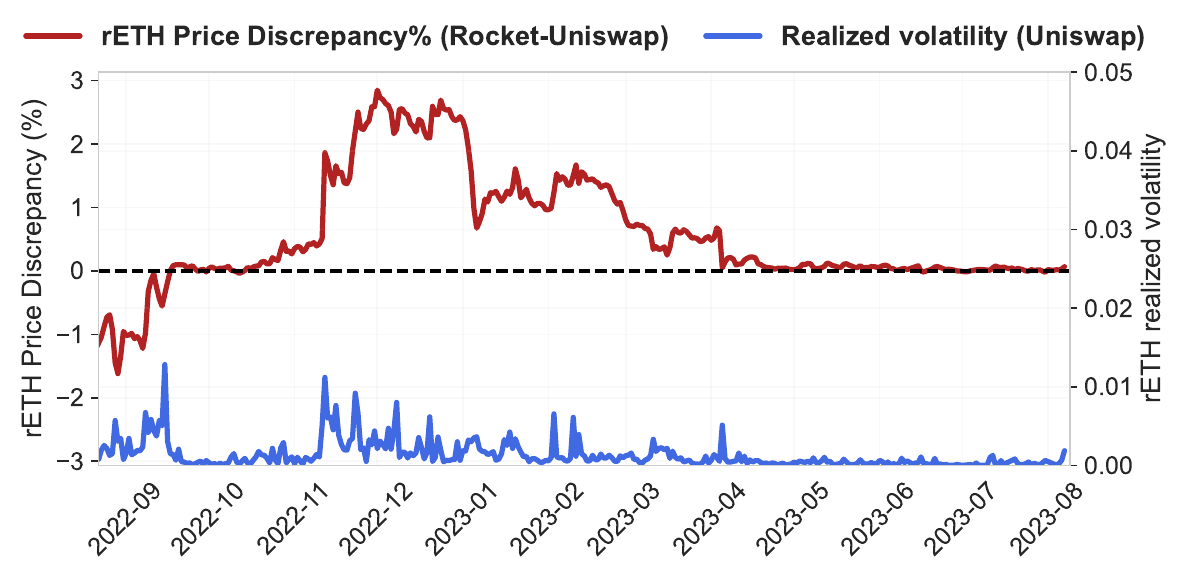}
    \caption{\rETH price discrepancy and realized volatility.}
    \label{fig:rETH_diff_vol}
\end{minipage}
\end{figure*}

\smallskip
Rebasing \LSDs adjust their token supply for reward distribution. Taking \stETH as an example, when users stake \ETH on Lido, they receive an equivalent 1:1 amount of \stETH. This indicates that the \stETH to \ETH price is fixed as $P_{\mathsf{stETH}}^{\mathsf{1st}}=1$ in the primary market. After obtaining \stETH, users can trade them in the secondary market through \DEXs such as \href{https://curve.fi/\#/ethereum/swap}{Curve}. While ideally, the secondary market price (\stETHPriceSecondary) should align with \stETHPricePrimary, in reality, a clear deviation between the two exists (see Fig.~\ref{fig: lido_wstETH_price_over_time}). To quantify such deviation, we crawl $10$-minute tick-level data of \stETHPricePrimary and \stETHPriceSecondary from Lido and Curve respectively. We calculate the Realized Volatility (RV)~\cite{andersen2009realized,mcaleer2008realized} of \stETHPriceSecondary on day $i$ using Eq.~\ref{eq:vol}. We observe an average $\mathsf{RV}_{\mathsf{stETH}}$ of $0.16\%$ and the maximum recorded $\mathsf{RV}_{\mathsf{stETH}}$ of $5\%$. We further calculate the price discrepancy between \stETHPricePrimary and \stETHPriceSecondary (see Eq.~\ref{eq:discrepancy}). We discover that on average, \stETHPriceSecondary is $0.83\%$ lower than \stETHPricePrimary. Moreover, we find that \stETHPriceSecondary deviates significantly from \stETHPricePrimary from May $7$ to May $16$, $2022$ due to the crash of \UST/\LUNA on the Terra network~\cite{briola2023anatomy, xiong2023leverage}. During this period, the price discrepancy widened to 6.9\%, reaching its all-time maximum~(see Fig.~\ref{fig:stETH_diff_vol}).



 \begin{figure*}[t]
    \centering
    \includegraphics[width=0.88\linewidth]{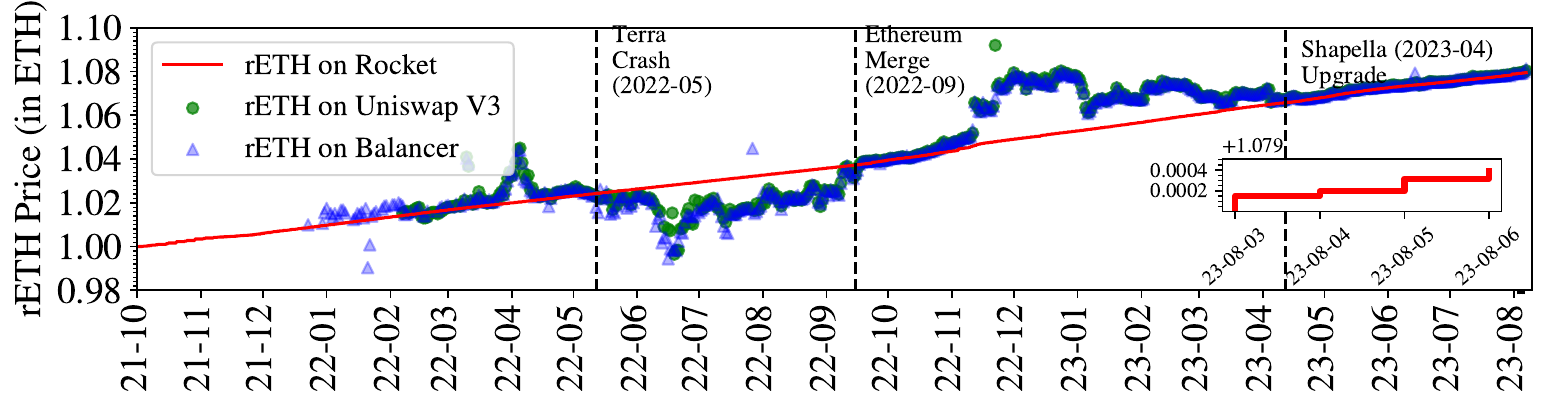}
    \caption{\rETH price on different platforms over time.}
    \label{fig:rETH_price_over_time}
\end{figure*}

Reward-bearing \LSDs, such as \rETH, accumulate rewards by adjusting their token value, leading to an increase in the \rETH price in the primary market (\rETHPricePrimary) over time. Consequently, the \rETH price in the secondary market (\rETHPriceSecondary) is anticipated to align with \rETHPricePrimary. To analyze the price behavior of \rETH, we gather $10$-minutes tick-level \rETHPricePrimary by querying the Rocket protocol contract. We also collect  \rETHPriceSecondary by querying Uniswap V3 and Balancer pool contracts respectively. We discover that \rETHPriceSecondary deviates from \rETHPricePrimary by an average of $0.22\%$. Furthermore, our data indicates that \rETHPriceSecondary is less volatile than \stETHPriceSecondary (see Fig.~\ref{fig:rETH_price_over_time}). This may be due to \rETH's constant supply, allowing smoother integration into \DEX designs as a reward-bearing token. Similar to \stETH, \rETH has also witnessed substantial volatility following the Terra crash, reaching a peak realized volatility of $3\%$~(see Fig.~\ref{fig:rETH_diff_vol}). Interestingly, \rETHPriceSecondary experienced a gradual rebound post-Merge and a notable upward trend subsequent to MakerDAO's introduction of the \rETH token in Nov 2022, suggesting a burgeoning rise in investor confidence.

\subsection{LSD Arbitrages}

The price discrepancy between the \LSD primary and secondary markets creates arbitrage opportunities. Capitalizing on these opportunities not only enables users to generate profits but also helps restore price equilibrium in different markets.

In the traditional financial market,  arbitrage~\cite{hausch1990arbitrage} exploits price discrepancies in various markets to secure profits without assuming any risk. Traders capitalize on temporary price differences, buying the asset at a lower price and selling it where it is higher. This approach is also applicable to \LSD arbitrage. For example, when \stETHPriceSecondary$>$\stETHPricePrimary, users can first stake \ETH on Lido to receive \stETH and sell \ETH immediately in the Curve pool to secure profits. However, in the context of \LSD arbitrage, this approach is not entirely risk-free due to \emph{(i)} the uncertainty to withdraw \ETH before Shapella, and \emph{(ii)} the potential slippage when trading on \DEXs.

\begin{figure*}[t]
\centering
\begin{minipage}{.49\textwidth}
  \centering
   \includegraphics[width=0.95\columnwidth]{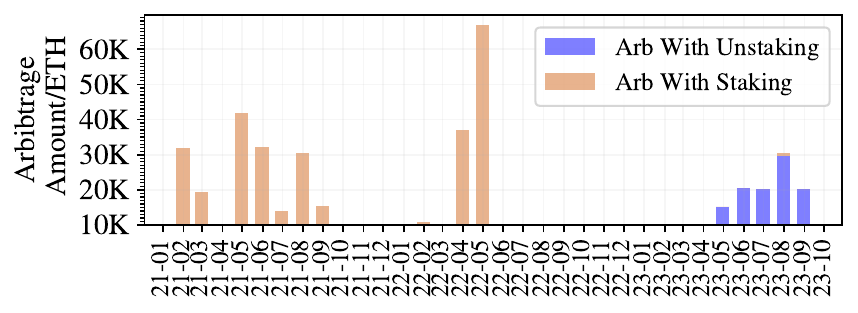}
    \caption{\stETH--\ETH arbitrage amount.}
    \label{fig: stETH_arbitrage_amount_over_time}
\end{minipage}%
\begin{minipage}{.49\textwidth}
  \centering
    \includegraphics[width=0.95\columnwidth]{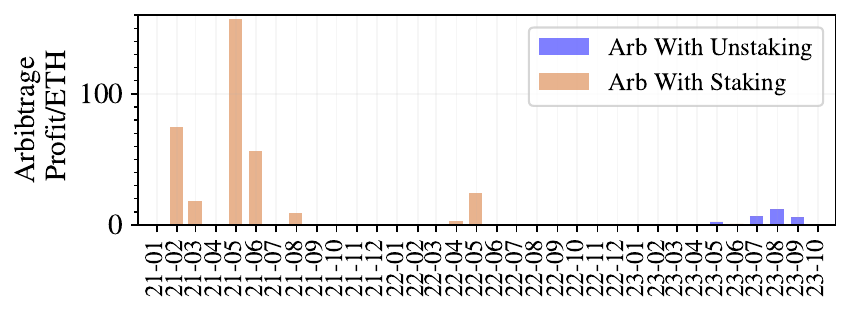}
    \caption{\stETH--\ETH arbitrage profit.}
    \label{fig: stETH_arbitrage_profit_over_time}
\end{minipage}
\end{figure*}

\begin{figure*}[t]
\centering
\begin{minipage}{.49\textwidth}
  \centering
   \includegraphics[width=0.95\columnwidth]{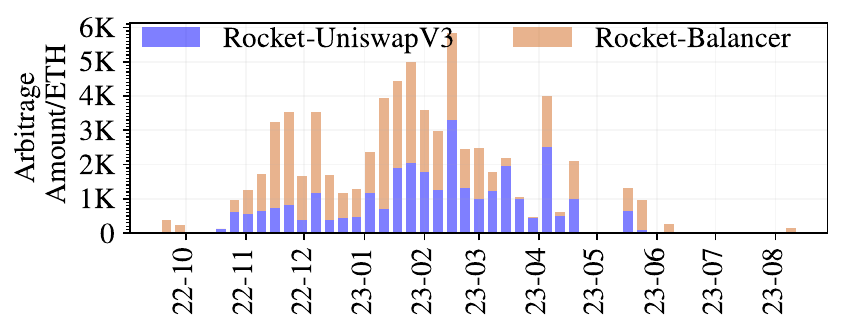}
    \caption{\rETH-\ETH arbitrage amount.}
    \label{fig: rETH_arbitrage_amount_over_time}
\end{minipage}%
\begin{minipage}{.49\textwidth}
  \centering
    \includegraphics[width=0.95\columnwidth]{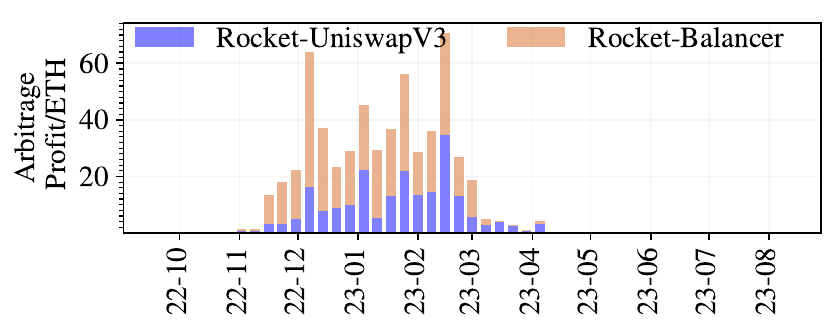}
    \caption{\rETH-\ETH arbitrage profit.}
    \label{fig: rETH_arbitrage_profit_over_time}
\end{minipage}
\end{figure*}

Note that users can also implement the arbitrage strategy when \stETHPriceSecondary$<$\stETHPricePrimary~(see Eq.~\ref{eq:arb_strategy}). In this scenario, users can initially exchange \ETH in the Curve pool for \stETH and subsequently redeem \stETH on Lido for \ETH. It is important to note that the arbitrage in this direction is only viable after the Shapella upgrade, as stakers are not allowed to redeem \stETH for \ETH before Shapella.
\begin{equation}\label{eq:arb_strategy}
    \begin{split}
        &\ETH \xrightarrow[\text{Lido}]{\text{Stake}} \stETH \xrightarrow[\text{Curve}]{\text{Swap}} \ETH {\small \text{, if } P_{\mathsf{stETH},t}^{\mathsf{2nd}}>P_{\mathsf{stETH},t}^{\mathsf{1st}}}\\
         &\ETH \xrightarrow[\text{Curve}]{\text{Swap}} \stETH \xrightarrow[\text{Lido}]{\text{Unstake}} \ETH {\small\text{, if } P_{\mathsf{stETH},t}^{\mathsf{2nd}}<P_{\mathsf{stETH},t}^{\mathsf{1st}}}\\
    \end{split}
\end{equation}

\subsection{Arbitrage Measurement}

\subsubsection{Arbitrages of Rebasing \LSDs} 

The price discrepancy, as illustrated in Fig.~\ref{fig: lido_wstETH_price_over_time}, between a rebased-based \LSD in the primary market and the secondary market creates opportunities for arbitrage. To systematically capture historical arbitrage events for \stETH--\ETH across Lido and Curve, we propose the following heuristics:

\begin{itemize}
    \item \emph{Arbitrage with Staking} (when $P_{\mathsf{stETH},t}^{\mathsf{2nd}}> P_{\mathsf{stETH},t}^{\mathsf{1st}}$): We crawl transactions where a user initially stakes \ETH on Lido to receive \stETH, followed by a subsequent swap of \stETH to \ETH on Curve. Notably, both the staking and swap events occur within the same transaction.
    
    \item \emph{Arbitrage with Unstaking} (when $P_{\mathsf{stETH},t}^{\mathsf{2nd}} < P_{\mathsf{stETH},t}^{\mathsf{1st}}$): We crawl the swap events where a user exchanges \ETH for \stETH on Curve after the Shapella upgrade. Subsequently, in separate transactions, the user unstakes \stETH on Lido to obtain \ETH.
\end{itemize}

\textbf{Arbitrage Amount and Profit.} We apply our heuristics to analyze \stETH--\ETH arbitrages across Lido and Curve from blocks $11{,}473{,}216$ (Dec~$17$,~$2020$) to $17{,}866{,}191$ (Aug $10$, $2023$). The results, depicted in Fig.~\ref{fig: stETH_arbitrage_amount_over_time} and Fig.~\ref{fig: stETH_arbitrage_profit_over_time}, reveal that $38$ addresses performed $400$ transactions for arbitrage with staking. These arbitrages accumulated a total of $200{,}741$~\ETH with an overall profit of $343$~\ETH. We observe that over $99\%$ of arbitrages with staking took place prior to Jun $2022$, a period when the price of \stETH on Curve occasionally exceeded $1$.

As for arbitrages in the reverse direction (e.g., with unstaking on Lido), we identify $42$ addresses participating in $55{,}617$~\ETH worth of \stETH--\ETH swaps after Apr $2023$, resulting in a total profit of $29$~\ETH.  In this case, following the swap of $m$~\ETH to $\frac{m}{P_{\mathsf{stETH}}^{\mathsf{2nd}}}$ \stETH on Curve, where the price is $1$~\stETH = $P_{\mathsf{stETH}}^{\mathsf{2nd}}$~\ETH (with $P_{\mathsf{stETH}}^{\mathsf{2nd}}<1$), arbitrageurs have the flexibility to execute the unstaking transaction at any time. This is due to the constant price of \stETH on Lido, always fixed at $1$, ensuring that arbitrageurs can secure a revenue of $(\frac{1}{P_{\mathsf{stETH}}^{\mathsf{2nd}}}-1)\cdot m$~\ETH after the unstaking process.

\begin{figure*}[t]
\centering
\begin{minipage}{.49\textwidth}
   \includegraphics[width =0.92\columnwidth]
  {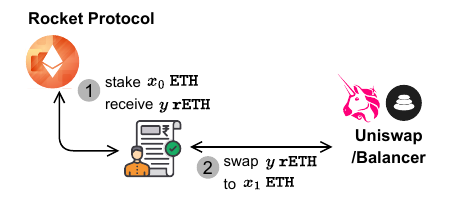}
  \label{fig:arbitrage}
\caption{Arbitrage without flash loans.}
\label{alg:arbitrage_without_flashloan}
\end{minipage}%
\hfill
\begin{minipage}{.49\textwidth}
   \includegraphics[width =0.9\columnwidth]
  {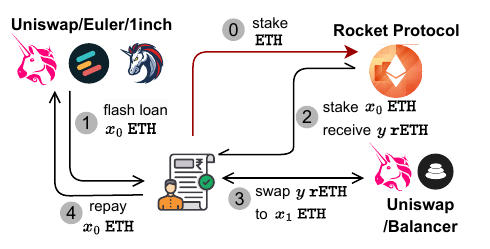}
  \caption{Arbitrage with flash loans.}
  \label{fig:arbitrage_with_flashloan}
\end{minipage}
\end{figure*}


\begin{figure}[t]
\centering
   \includegraphics[width =0.95\columnwidth]
{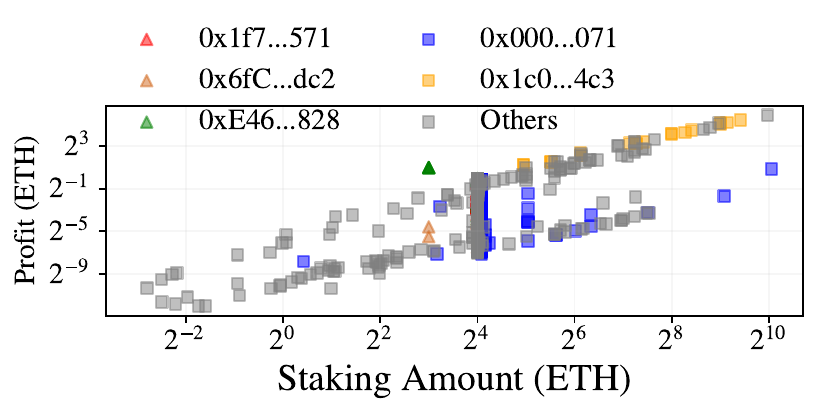}
\caption{Arbitrage profit and staking.}
\label{fig:rETH_arbitrage_profit_over_staking_amount}
\end{figure}

\vspace{+2mm}
\subsubsection{Arbitrages of Reward-bearing \LSDs}
The price of a reward-bearing LSD experiences periodic increments to reflect the accumulative staking reward within the network. As shown in Fig.~\ref{fig:rETH_price_over_time}, the price of \rETH on Rocket Pool witnesses a daily increase whenever the oracle updates the beacon's reward allocations garnered by the validators. Nonetheless, the secondary market price does not consistently align with the primary market price. Such price discrepancies across different platforms can create opportunities for arbitrage.


\textbf{Arbitrage Amount and Profit.} We investigate the arbitrage opportunities involving \rETH on Rocket-Balancer and Rocket-Uniswap V3 from Sep $30$,~$2021$, to Aug $10$,~$2023$. Our analysis reveals that $373$ addresses executed $3{,}735$ arbitrage transactions during this period, resulting in a cumulative exchange of $69{,}270.7$~\ETH and an overall profit of $577.1$~\ETH. The distributions of \rETH-\ETH arbitrage amounts and profits over time are shown in Fig.~\ref{fig: rETH_arbitrage_amount_over_time} and~\ref{fig: rETH_arbitrage_profit_over_time} respectively.  Interestingly, over $99\%$ of the arbitrages occurred prior to May $2023$, aligning with the historical price volatility of \rETH. This pattern is evident in Fig.~\ref{fig:rETH_price_over_time}, which illustrates the nearly identical price trajectories of \rETH on Uniswap V3, Balancer, and the Rocket protocol after May $2023$.



\vspace{+2mm}

\textbf{Arbitrage Strategies Analysis.} Interestingly, we observe that the $3{,}811$ arbitrage transactions were executed by interacting with $35$ distinct contract addresses (see Table~\ref{tab:rocket_arbitrages} in Appendix~\ref{sec:app-addtional-empirical}). Out of these, $10$ contract addresses, also known as arbitrage bots, were invoked by multiple arbitrageurs. For instance, \href{https://etherscan.io/address/0x1f7e55F2e907dDce8074b916f94F62C7e8A18571}{0x1f7...571} was utilized by $329$ arbitrageurs to initiate $1{,}903$ arbitrage transactions from Nov $7$, $2022$ to Mar $17$, $2023$, generating a cumulative profit of $1{,}257.8$~\ETH.

After manually assessing these $35$ contract addresses, we discover that $4$ of them feature publicly accessible code, with $2$ written in Vyper and $2$ in Solidity. After a thorough analysis of their transactions and code, we compile their arbitrage strategy particulars, which can be grouped into two categories:

\noindent
\begin{itemize}[leftmargin=*]
    \item \emph{Arbitrage without flash loan} (see Fig.~\ref{alg:arbitrage_without_flashloan}): Upon observing the price disparities (\rETHPricePrimary<\rETHPriceSecondary) of \rETH-\ETH token pair on Rocket protocol ($1~\rETH = P_{\mathsf{rETH}}^{\mathsf{1st}}~\ETH$) and \DEXs ($1~\rETH = P_{\mathsf{rETH}}^{\mathsf{2nd}}~\ETH$), an arbitrageur performs the following process: \emph{(i)} stakes $x_0$~\ETH on Rocket protocol to receive $y =x_0/P_{\mathsf{rETH}}^{\mathsf{1st}}$~\rETH; \emph{(ii)} swaps $y$~\rETH to $x_1 = y\cdot P_{\mathsf{rETH}}^{\mathsf{2nd}}$~\ETH on \DEXs. All these steps occur on a single transaction $\tx_{arb}$. The final profit is $x_1 - x_0 - cost(\tx_{arb})$, where the transaction cost $cost(\tx_{arb})$ includes the gas fees and the fee used to bribe validators.

    \vspace{+1mm}
    \item \emph{Arbitrage with flash loan} (see Fig.~\ref{fig:arbitrage_with_flashloan}): Upon observing the price disparities of \rETH-\ETH token pair, an arbitrageur performs the following process: \emph{(i)} borrows $x_0$~\ETH from the DeFi platforms supporting flash loans~\cite{wang2021towards} (e.g., Uniswap, Euler, 1inch); \emph{(ii)} stakes $x_0$~\ETH on Rocket protocol to receive $y = x_0/P_{\mathsf{rETH}}^{\mathsf{1st}}$~\rETH; \emph{(iii)} swaps $y$~\rETH to $x_1 = y\cdot P_{\mathsf{rETH}}^{\mathsf{2nd}}$~\ETH on \DEXs; \emph{(iii)} repays the $x_0$~\ETH flash loans. All these steps occur on a single transaction $\tx_{arb}$. The final profit of the arbitrage is $x_1 - x_0 - cost(\tx_{arb})$.
\end{itemize}

\textbf{Is there a barrier to engaging in \rETH-\ETH arbitrage with flash loan?} We identify four \rETH-\ETH arbitrage bots whose code is publicly accessible. Interestingly, due to the code transparency, any user who witnesses a profitable on-chain arbitrage transaction $\tx_{arb}$ can replicate the corresponding arbitrageur's strategy. This involves creating a transaction with identical input data as that of $\tx_{arb}$ to invoke the arbitrage bot. By leveraging flash loans, the user can execute this action without the necessity of transferring any \ETH to the bots, merely incurring the transaction cost $cost(\tx_{arb})$.


However, this seemingly straightforward ``copy-paste'' arbitrage approach does not yield practical results. This is because, in real-world scenarios, prior to executing an arbitrage transaction with flash loan, an arbitrageur needs to either stake \ETH or await the staking of \ETH by other users within the Rocket protocol (e.g., see the transactions with index $51$ and $52$ in block $17{,}073{,}005$). Such actions trigger changes in the Rocket protocol which enables further staking\footnote{See \href{https://github.com/rocket-pool/rocketpool/blob/master/contracts/contract/deposit/RocketDepositPool.sol\#L113}{Line 113 of RocketDepositPool.sol}}. Our findings reveal that at least $2{,}037$ successful flash loan-based \rETH-\ETH arbitrage transaction $\tx_{arb}$ consistently occurs just after a staking transaction $\tx_{stake}$ issued by the same arbitrageur. In other words, $\tx_{arb}$ and $\tx_{stake}$ are placed in the same block and possess consecutive transaction indexes. Our empirical findings indicate that arbitrageurs encounter a substantial entry barrier, necessitating an average stake of $16$~\ETH as a prerequisite for participating in \rETH arbitrage.

Fig.~\ref{fig:rETH_arbitrage_profit_over_staking_amount} illustrates the distribution of profits from \rETH-\ETH arbitrage in relation to the staked amount of \ETH. Notably, upon excluding arbitrages involving a staking amount of 16~\ETH, a clear linear increase in profits over the staked amount becomes evident, characterized by two distinct slopes. For instance, the arbitrage bot with the address \href{https://etherscan.io/address/0x1c073d5045b1aBb6924D5f0f8B2F667b1653a4c3}{0x1c0...4c3} generated a total profit of $292.8$~\ETH, staking a cumulative amount of $4{,}418.8$~\ETH.
We also analyze the distribution of arbitrages with a staking amount of $16$~\ETH (see Fig.~\ref{fig:rETH_arbitrage_profit_distribution_staking_16eth}), and find that they yielded $0.49\pm 0.33$~\ETH on average.

\section{Liquidity Provision with \LSDs}



 In this section, our focus shifts to analyzing \LP's financial incentives for engaging in liquidity provision with \LSDs.

After acquiring \LSDs in the primary market, users face two options. They can either opt to retain the \LSDs until the Shapella upgrade, at which point they can unstake the \LSDs to withdraw \ETH and receive the associated staking rewards. Alternatively, users may leverage their \LSDs to explore wider financial opportunities in the secondary market, such as participating in liquidity provision for \DEX pools.

\subsection{PNL of LSD Liquidity Provision}

At the time of writing, \href{https://curve.fi/#/ethereum/swap}{Curve} stands out as the dominant \DEX for \stETH, whereas \href{https://app.uniswap.org/swap}{Uniswap} and \href{https://app.balancer.fi/\#/ethereum/pool/0x1e19cf2d73a72ef1332c882f20534b6519be0276000200000000000000000112}{Balancer} take the lead as the most liquid \DEXs for \rETH. Fig.~\ref{fig:lsd_trading_over_time} shows the distribution of \LSD trading volume on Curve, Uniswap V3 and Balancer over time. We find that the accumulated trading volume of \stETH on Curve from Feb 2021 to Aug 2023 is $11{.}2$m~\ETH, while the cumulative trading volume of \rETH is $775$k~\ETH since the inception of the \rETH-\ETH pools on Balancer and Uniswap V3 in Dec 2021 and Feb 2022. This result suggests that \stETH experiences more active trading in the secondary market than \rETH.

\smallskip
In light of the increasing trend of \LSD liquidity provision, we aim to examine the \PNL experienced by \LPs and assess how these \PNLs differ among \LSDs with distinct token mechanisms. Consider a user intending to add an initial amount of $q_{x,t_0}$ \LSD and $q_{y,t_0}$ \ETH to a \LSD-ETH liquidity pool at time $t_0$. Given the spot price of $p_{x,t_0}$, 
the initial portfolio value in \ETH is $V_{t_0}^{\text{LP}} = q_{y,t_0} + q_{x,t_0}\cdot p_{x,t_0}$. 
While users can gain financial benefits through liquidity provision, they may also encounter losses in their portfolio value due to price changes. In particular, the user's \PNL originates from four sources: \emph{(i)} the swap fees earned through liquidity provision; \emph{(ii)} the accumulated staking reward; \emph{(iii)} the change in portfolio value due to price volatility; \emph{(iv)} the transaction fees associated with adding and removing liquidity.

\smallskip
Suppose that at time $t_1$, the \LSD price changes to $p_{x,t_1}$ and the user can withdraw $q_{x,t_1}^{\text{LP}}$ \LSD and $q_{y,t_1}^{\text{LP}}$ \ETH. User's portfolio value changes from $V_{t_0}^{\text{LP}}$ to $V_{t_1}^{\text{LP}}=q_{y,t_1}^{\text{LP}} + q_{x,t_1}^{\text{LP}}\cdot p_{x,t_1}$. Note that the earned swap fees and accumulated staking rewards are already reflected in $V_{t_1}^{\text{LP}}$ when \LP removes liquidity from the pool. Specifically, if the user supplies \stETH for liquidity provision, the rewards will be manifested in the withdrawal amount of \stETH (i.e., $q_{x,t_1}^{\text{LP}}$). If the user provides \rETH to the liquidity pool, the rewards will be expressed through changes in \rETH price over time. Therefore, we can calculate the liquidity provision \PNL using Eq.~\ref{eq:PRN_LP}, where function $f(\cdot)$ converts the periodical rate of return to \APR. 
\begin{equation}\label{eq:PRN_LP}
    {\small
    \begin{split}
    &V_{t_0}^{\text{LP}}=q_{y,t_0} + q_{x,t_0}\cdot p_{x,t_0}\text{, }V_{t_1}^{\text{LP}}=q_{y,t_1}^{\text{LP}} + q_{x,t_1}^{\text{LP}}\cdot p_{x,t_1}\\
    &\text{PNL}_{(t_0,t_1)}^{\text{LP}} = V_{t_1}^{\text{LP}} - V_{t_0}^{\text{LP}}\text{, } \text{APR}_{(t_0,t_1)}^{\text{LP}}=  f(\frac{\text{PNL}_{(t_0,t_1)}^{\text{LP}}}{V_{t_0}^{\text{LP}}})\\
    \end{split}
    }
\end{equation}

\begin{figure*}[t]
\centering
\subfigure[\ETH-\stETH on Curve.]{
\includegraphics[width=0.312\linewidth]{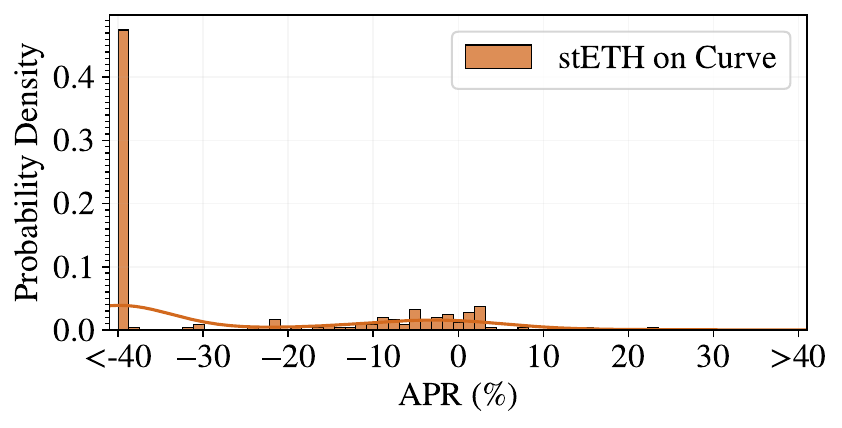}
\label{fig: lsd_realized_profit_apr_stETH_curve_w_tx_fee_distribution}
}%
\subfigure[\ETH-\rETH on Uniswap V3.]{
\includegraphics[width=0.312\linewidth]{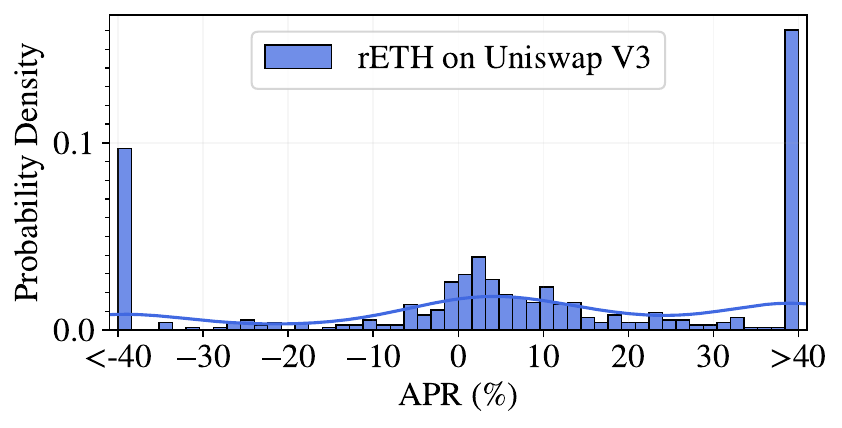}
\label{fig: lsd_realized_profit_apr_rETH_uniswapv3_w_tx_fee_distribution}
}%
\hfill
\subfigure[\ETH-\rETH on Balancer.]{
\includegraphics[width=0.312\linewidth]{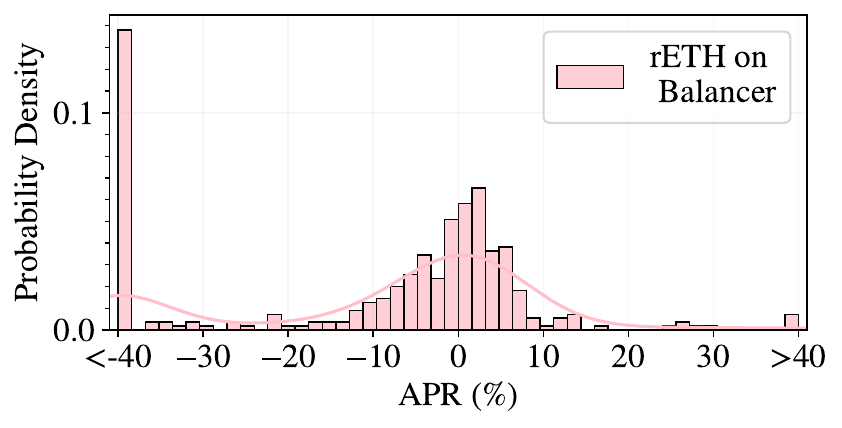}
\label{fig: lsd_realized_profit_apr_rETH_balancer_w_tx_fee_distribution}
}%
\centering
\caption{Distribution of net APRs for \LSD liquidity provision on \DEXs (including transaction fees).}
\label{fig:lp-apr-with-tx-fee}
\end{figure*}

\begin{figure*}[t]
\centering
\subfigure[\ETH-\stETH On Curve.]{
\includegraphics[width=0.312\linewidth]{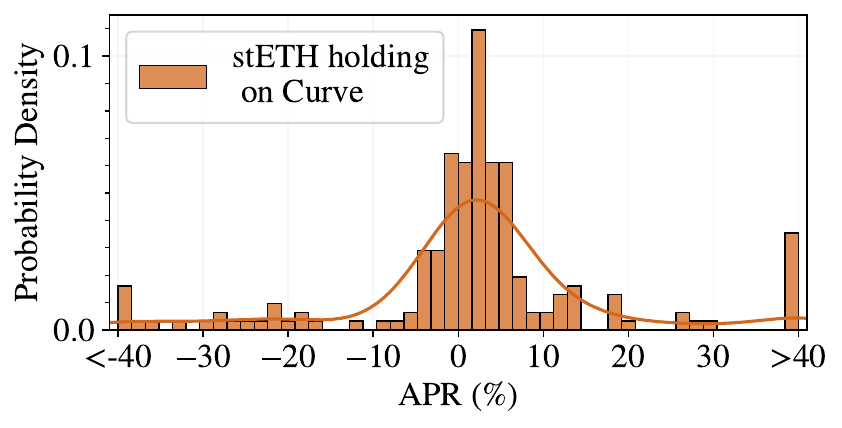}
\label{fig: lsd_holding_realized_profit_apr_stETH_curve_distribution}
}%
\subfigure[\ETH-\rETH On Uniswap V3.]{
\includegraphics[width=0.312\linewidth]{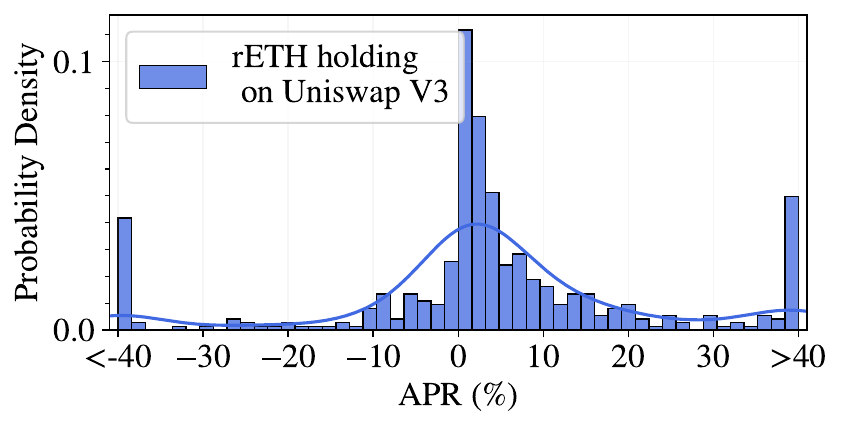}
\label{fig: lsd_holding_realized_profit_apr_rETH_uniswapv3_distribution}
}%
\hfill
\subfigure[\ETH-\rETH On Balancer.]{
\includegraphics[width=0.312\linewidth]{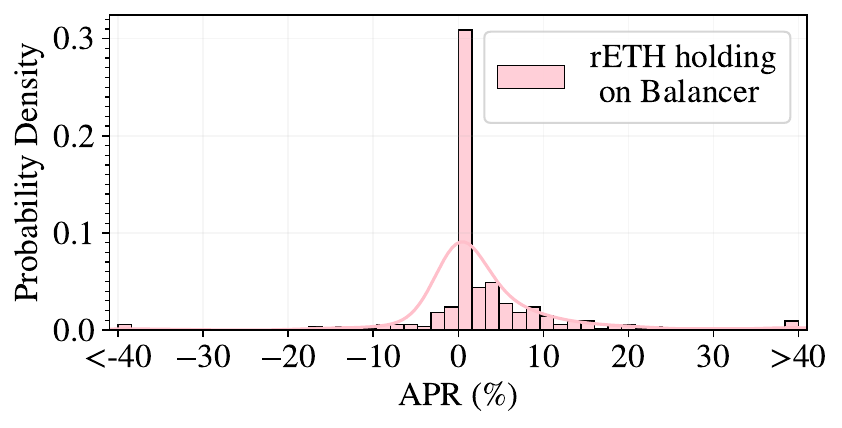}
\label{fig: lsd_holding_realized_profit_apr_rETH_balancer_distribution}
}%
\centering
\caption{Distribution of net APRs for simply holding \LSDs.}
\label{fig: actual-holding-apr}
\end{figure*}

\begin{figure*}[t]
\centering
\subfigure[\ETH-\stETH on Curve.]{
\includegraphics[width=0.312\linewidth]{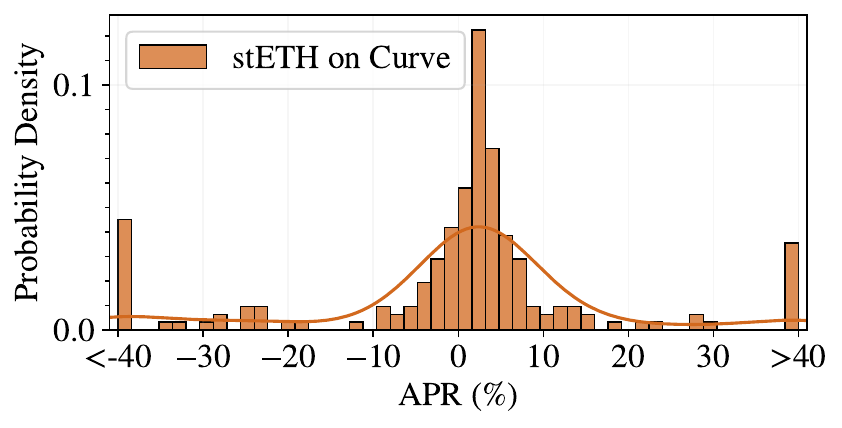}
\label{fig: lsd_realized_profit_apr_stETH_curve_wo_tx_fee_distribution}
}%
\subfigure[\ETH-\rETH on Uniswap V3.]{
\includegraphics[width=0.312\linewidth]{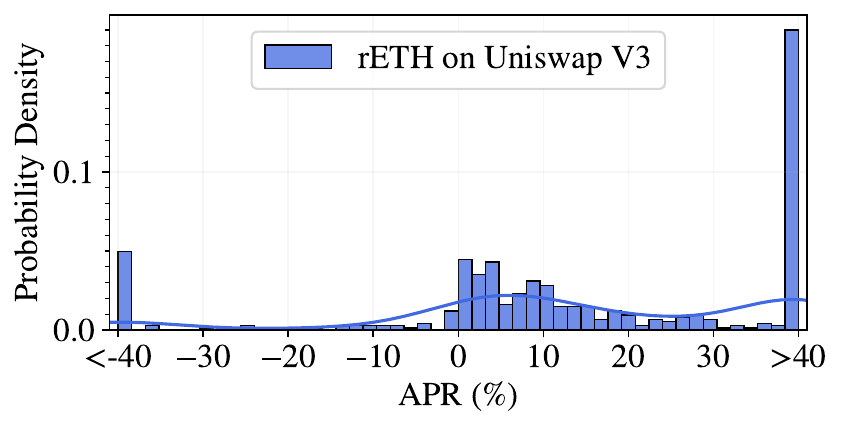}
\label{fig: lsd_realized_profit_apr_rETH_uniswapv3_wo_tx_fee_distribution}
}%
\hfill
\subfigure[\ETH-\rETH on Balancer.]{
\includegraphics[width=0.312\linewidth]{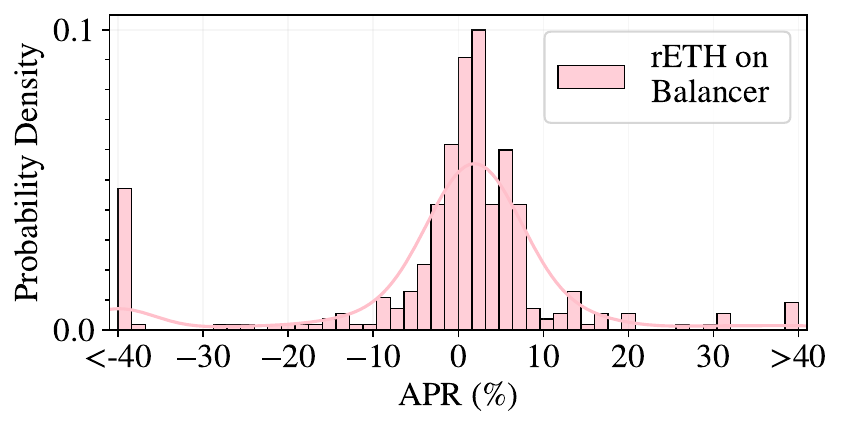}
\label{fig: lsd_realized_profit_apr_rETH_balancer_wo_tx_fee_distribution}
}%
\centering
\caption{Distribution of APRs for \LSD liquidity provision on \DEXs (excluding transaction fees).}
\label{fig:lp-apr-wo-tx-fee}
\end{figure*}

\textbf{Empirical Analysis.} 
To analyze the \APR obtained by users providing \LSDs as liquidity in \DEXs, we explore transactions that add and remove liquidity in \LSDs on Curve, Uniswap V3, and Balancer up to block $17{,}866{,}191$ (Aug 7th, 2023). Our investigation identifies $34{,}682$ \texttt{AddLiquidity} and $11{,}985$ \texttt{RemoveLiquidity} events on the \href{https://etherscan.io/address/0xdc24316b9ae028f1497c275eb9192a3ea0f67022}{Curve \stETH pool}, $878$ \texttt{Mint} and $1{,}328$ \texttt{collect} events on the \href{https://etherscan.io/address/0xa4e0faA58465A2D369aa21B3e42d43374c6F9613}{Uniswap V3 \rETH pool}, as well as $3{,}474$ \texttt{PoolBalanceChanged} events on the \href{https://etherscan.io/address/0x1e19cf2d73a72ef1332c882f20534b6519be0276}{Balancer \rETH pool}. Within these events, we identify $194$, $464$, and $344$ liquidity positions in which users have withdrawn all of their \LSDs from the Curve, Uniswap V3, and Balancer, respectively. We use Equation~\ref{eq:PRN_LP} to calculate the actual \APR of the $1{,}002$ liquidity provision positions.

\smallskip
Figures~\ref{fig:lp-apr-wo-tx-fee} and~\ref{fig:lp-apr-with-tx-fee} show the probability distribution of \LSD liquidity provision \APR on the three \DEXs, excluding and including transaction fees respectively. When making horizontal comparisons, we discover that the \LSD liquidity provision \APR is likely to be influenced by the underlying token mechanism. Specifically, when transaction fees are not considered, we find that the \APRs for $71.7\%$ of the detected \stETH liquidity provision positions are concentrated around $[-0.1,0.1]$, whereas only $36.8\%$ of \rETH liquidity provision positions exhibit \APRs fall within the same range. In fact, $21.3\%$  of the \rETH positions achieve an \APR greater than $1$. When making a vertical comparison to consider transaction fees, the difference is more pronounced. Our results show that $85.3\%$ of \stETH liquidity provision positions experience a negative \APR, whereas the majority ($69.8\%$) of \rETH liquidity provision positions achieved a positive \APR.

\xihan{
We also note that the performance of liquidity provision for the same \LSD varies among different \DEX pools. When comparing \rETH liquidity provision on Uniswap V3 and Balancer, we observe that users on Uniswap V3 are more likely to achieve higher \APRs. This difference may be attributed to the concentrated liquidity design of Uniswap V3~\cite{xiong2023demystifying}, allowing users to enhance capital efficiency through liquidity provision.
}

\subsection{\LSD Liquidity Provision vs Holding Strategy}

Users are incentivized to supply \LSDs for liquidity provision instead of simply holding them due to the prospect of increased profitability. Consequently, we undertake an empirical study to analyze the \PNLs associated with these two strategies. 
\begin{equation}\label{eq:pnl_hold}
    \small{
    \begin{split}
         & V_{t_0}^{\text{Hold}}=q_{y,t_0} + q_{x,t_0}\cdot p_{x,t_0}\text{, }V_{t_1}^{\text{Hold}}=q_{y,t_0}+ q_{x,t_1}^{\text{Hold}}\cdot p_{x,t_1} \\
        &\text{PNL}_{(t_0,t_1)}^{\text{Hold}} = V_{t_1}^{\text{Hold}} - V_{t_0}^{\text{Hold}}\text{, } \text{APR}_{(t_0,t_1)}^{\text{Hold}}=  f(\frac{\text{PNL}_{(t_0,t_1)}^{\text{Hold}}}{V_{t_0}^{\text{Hold}}})
    \end{split}
    }
\end{equation}

The user has the opportunity to accumulate \LSD staking rewards through both strategies, but it's crucial to note that the staking reward for liquidity provision differs from that of holding \LSDs, primarily due to the impact of swaps within the liquidity pool. Furthermore, \LSDs with different token mechanisms implement varied approaches to distribute staking rewards. \stETH adopts a rebasing model,  while \rETH utilizes a reward-bearing model, changing the price of \rETH at time $t_1$ for reward distribution. Hence, the quantity of \ETH remains unchanged at time $t_1$, whereas the price and quantity of \LSDs change based on the underlying token mechanisms. The \PNL of the holding strategy can be derived using Eq.~\ref{eq:pnl_hold}.


\textbf{Empirical Analysis.} For the $1{,}002$ identified liquidity positions, we simulate their portfolio value at $t_1$ as if users had chosen to hold the corresponding \LSD. As \rETH is based on a reward-bearing model (see Fig.~\ref{fig: token_models}), the holding amounts at $t_0$ and $t_1$ are the same, i.e., $q_{\rETH,t_0} \equiv q_{\rETH,t_1}$. For \stETH, which is based on a rebasing model, we query the underlying Lido contracts to obtain the values of $\mathsf{totalEther}(t)$ and $\mathsf{totalShares}(t)$ at a given time and apply Eq.~\ref{eq:rebasing} to compute the newly holding amount of \stETH at $t_1$. We finally calculate the holding \PNL and \APR by using Eq.~\ref{eq:pnl_hold}.

In total, we find that $660 \thinspace (66\%)$ identified liquidity positions can achieve a higher \APR by holding the \LSDs rather than supplying them to \DEXs. Specifically, as depicted in Fig.~\ref{fig: actual-holding-apr}, the distribution of the holding \APR for both \stETH and \rETH closely approximates a normal distribution, with average values ranging between $-14\%$ and $14\% $. By comparing the net holding \APR in Fig.~\ref{fig: lsd_holding_realized_profit_apr_stETH_curve_distribution} and the net \APR of \stETH liquidity provision on Curve in Fig.~\ref{fig: lsd_realized_profit_apr_stETH_curve_w_tx_fee_distribution}, it becomes apparent that users are likely to achieve a positive \APR if they opt to hold \stETH or \rETH in their wallets. This is primarily because they can avoid incurring transaction fees associated with adding and removing liquidity on Curve. Conversely, users who choose to provide \rETH on Uniswap V3 may have a higher probability (see the bars of ``$>40$'' in Figures~\ref{fig: lsd_realized_profit_apr_rETH_uniswapv3_w_tx_fee_distribution} and ~\ref{fig: lsd_holding_realized_profit_apr_rETH_uniswapv3_distribution}) of attaining a substantial \APR compared to holding \rETH, but they also face an increased risk of potential asset losses (see the bars of ``$<-40$'' in Figures~\ref{fig: lsd_realized_profit_apr_rETH_uniswapv3_w_tx_fee_distribution} and ~\ref{fig: lsd_holding_realized_profit_apr_rETH_uniswapv3_distribution}).

\section{Conclusion}

This paper provides an empirical study on \LSDs. We first systematize the existing \LSD token mechanisms and their functionality in the distribution of staking rewards. Then we quantify price discrepancies for \stETH and \rETH across primary and secondary markets. We identify the historical arbitrages associated with \LSDs, analyze strategies adopted by arbitrageurs, and shed light on potential entry barriers for \LSD arbitrages, particularly with \rETH. In addition, we measure the \APR achieved by \LPs who supply \LSDs to \DEXs, revealing that the yield from liquidity provision can be influenced by various factors, such as the underlying token mechanisms of \LSDs, \DEX pool designs, and transaction fees. Moreover, our analysis indicates that liquidity provision with \LSDs does not always result in higher returns compared to simply holding the assets. We hope our study can inspire further research into the analysis and design of \LSDs.


\bibliographystyle{ieeetr}
\bibliography{references.bib}

\begin{thebibliography}{10}

\bibitem{bitcoin}
S.~Nakamoto, ``Bitcoin: A peer-to-peer electronic cash system,'' 2008.
\newblock Available at: \url{https://bitcoin.org/bitcoin.pdf}.

\bibitem{wood2014ethereum}
G.~Wood, ``Ethereum: A secure decentralised generalised transaction ledger,''
  {\em Ethereum project yellow paper}, vol.~151, pp.~1--32, 2014.

\bibitem{daian2019snow}
P.~Daian, R.~Pass, and E.~Shi, ``Snow white: Robustly reconfigurable consensus
  and applications to provably secure proof of stake,'' in {\em FC},
  pp.~23--41, Springer, 2019.

\bibitem{buterin2020combining}
V.~Buterin, D.~Hernandez, T.~Kamphefner, K.~Pham, Z.~Qiao, D.~Ryan, J.~Sin,
  Y.~Wang, and Y.~X. Zhang, ``Combining {GHOST} and casper,'' {\em arXiv
  preprint arXiv:2003.03052}, 2020.

\bibitem{gavzi2019proof}
P.~Ga{\v{z}}i, A.~Kiayias, and D.~Zindros, ``Proof-of-stake sidechains,'' in
  {\em IEEE S\&P}, pp.~139--156, IEEE, 2019.

\bibitem{kiayias2017ouroboros}
A.~Kiayias, A.~Russell, B.~David, and R.~Oliynykov, ``Ouroboros: A provably
  secure proof-of-stake blockchain protocol,'' in {\em Annual International
  Cryptology Conference (CRYPTO)}, pp.~357--388, Springer, 2017.

\bibitem{zhang2023rationally}
Y.~Zhang, Q.~Wang, S.~Chen, and C.~Wang, ``How to rationally select your
  delegatee in {PoS},'' {\em arXiv preprint arXiv:2310.08895}, 2023.

\bibitem{scharnowski2023economics}
S.~Scharnowski and H.~Jahanshahloo, ``The economics of liquid staking
  derivatives: Basis determinants and price discovery,'' {\em Available at SSRN
  4180341}, 2023.

\bibitem{xiong2023leverage}
X.~Xiong, Z.~Wang, X.~Chen, W.~Knottenbelt, and M.~Huth, ``Leverage staking
  with liquid staking derivatives ({LSD}s): Opportunities and risks.''
  Cryptology ePrint Archive, Paper 2023/1842, 2023.

\bibitem{tzinas2023principal}
A.~Tzinas and D.~Zindros, ``The principal--agent problem in liquid staking,''
  in {\em International Conference on Financial Cryptography and Data Security
  (FC)}, pp.~456--469, Springer, 2023.

\bibitem{werner2022sok}
S.~Werner, D.~Perez, L.~Gudgeon, A.~Klages-Mundt, D.~Harz, and W.~Knottenbelt,
  ``Sok: Decentralized finance ({DeFi}),'' in {\em AFT}, pp.~30--46, 2022.

\bibitem{jiang2023decentralized}
E.~Jiang, B.~Qin, Q.~Wang, Z.~Wang, Q.~Wu, J.~Weng, X.~Li, C.~Wang, Y.~Ding,
  and Y.~Zhang, ``Decentralized finance ({DeFi}): A survey,'' {\em arXiv
  preprint arXiv:2308.05282}, 2023.

\bibitem{grandjean2023ethereum}
D.~Grandjean, L.~Heimbach, and R.~Wattenhofer, ``Ethereum proof-of-stake
  consensus layer: Participation and decentralization,'' {\em International
  Conference on Financial Cryptography and Data Security Workshops
  (CoDecFin@FC)}, 2023.

\bibitem{schwarz2022three}
C.~Schwarz-Schilling, J.~Neu, B.~Monnot, A.~Asgaonkar, E.~N. Tas, and D.~Tse,
  ``Three attacks on proof-of-stake {E}thereum,'' in {\em FC}, pp.~560--576,
  Springer, 2022.

\bibitem{agrawal2022proofs}
S.~Agrawal, J.~Neu, E.~N. Tas, and D.~Zindros, ``Proofs of proof-of-stake with
  sublinear complexity,'' in {\em AFT}, 2023.

\bibitem{tang2023transaction}
W.~Tang and D.~D. Yao, ``Transaction fee mechanism for proof-of-stake
  protocol,'' {\em arXiv preprint arXiv:2308.13881}, 2023.

\bibitem{kapengut2023event}
E.~Kapengut and B.~Mizrach, ``An event study of the ethereum transition to
  proof-of-stake,'' {\em Commodities}, vol.~2, no.~2, pp.~96--110, 2023.

\bibitem{EthMerge2022}
``The merge,'' 2023.
\newblock Available at: \url{https://ethereum.org/en/roadmap/merge/}.

\bibitem{EIP-4895}
``Eip-4895: Beacon chain push withdrawals as operations,'' 2022.
\newblock Available at: \url{https://eips.ethereum.org/EIPS/eip-4895}.

\bibitem{cassez2022formal}
F.~Cassez, J.~Fuller, and A.~Asgaonkar, ``Formal verification of the {E}thereum
  2.0 beacon chain,'' in {\em TACAS}, pp.~167--182, Springer, 2022.

\bibitem{andersen2009realized}
T.~G. Andersen and T.~Ter{\"a}svirta, ``Realized volatility,'' in {\em Handbook
  of Financial Time Series}, pp.~555--575, Springer, 2009.

\bibitem{mcaleer2008realized}
M.~McAleer and M.~C. Medeiros, ``Realized volatility: A review,'' {\em
  Econometric Reviews}, vol.~27, no.~1-3, pp.~10--45, 2008.

\bibitem{briola2023anatomy}
A.~Briola, D.~Vidal-Tom{\'a}s, Y.~Wang, and T.~Aste, ``Anatomy of a
  stablecoin’s failure: The terra-luna case,'' {\em Finance Research
  Letters}, vol.~51, p.~103358, 2023.

\bibitem{hausch1990arbitrage}
D.~B. Hausch and W.~T. Ziemba, ``Arbitrage strategies for cross-track betting
  on major horse races,'' {\em Journal of Business}, pp.~61--78, 1990.

\bibitem{wang2021towards}
D.~Wang, S.~Wu, Z.~Lin, L.~Wu, X.~Yuan, Y.~Zhou, H.~Wang, and K.~Ren, ``Towards
  a first step to understand flash loan and its applications in defi
  ecosystem,'' in {\em Proceedings of International Workshop on Security in
  Blockchain and Cloud Computing (SBC@AsiaCCS)}, pp.~23--28, 2021.

\bibitem{xiong2023demystifying}
X.~Xiong, Z.~Wang, W.~Knottenbelt, and M.~Huth, ``Demystifying just-in-time
  ({JIT}) liquidity attacks on {U}niswap {V3},'' {\em Conference on Blockchain
  Research \& Applications for Innovative Networks and Services (BRAINS)},
  pp.~1--8, 2023.

\end{thebibliography}

\appendix


\section*{Supplementary Information} \label{sec:app-addtional-empirical}

\begin{equation}\label{eq:vol}
  {\small
    \mathsf{RV}_{\mathsf{stETH},i} = \sqrt{\sum_{j=1}^{144} (\mathsf{ln}\frac{P_{\mathsf{stETH},tick_{i,j+1}}^{\mathsf{2nd}}}{P_{\mathsf{stETH},tick_{i,j}}^{\mathsf{2nd}}})^2}
    }
\end{equation}

\begin{equation}\label{eq:discrepancy}
  {\small
    \mathsf{PD}_{\mathsf{stETH},i} = \mathsf{avg} (\sum_{j=1}^{144} \frac{P_{\mathsf{stETH},tick_{i,j}}^{\mathsf{2nd}}-P_{\mathsf{stETH},tick_{i,j}}^{\mathsf{1st}}}{P_{\mathsf{stETH},tick_{i,j}}^{\mathsf{1st}}})
    }
\end{equation}



Eq.~\ref{eq:vol} and~\ref{eq:discrepancy} are used to calculate the price discrepancy between the \LSD primary market and secondary market.

\begin{figure}[t]
\centering
\begin{minipage}{.42\textwidth}
   \includegraphics[width =\columnwidth]
{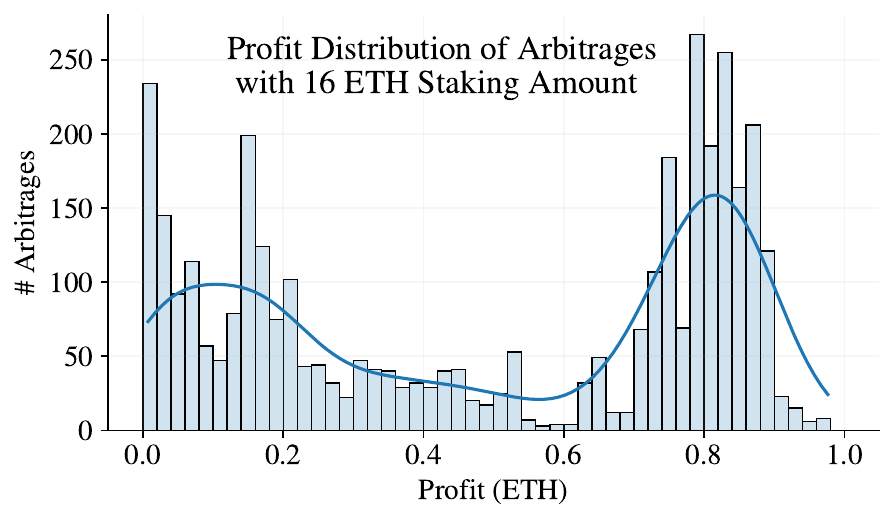}
  \caption{Arbitrage profits with staking
  $16$~\ETH.}
\label{fig:rETH_arbitrage_profit_distribution_staking_16eth}
\end{minipage}%

\begin{minipage}{.42\textwidth}
   \includegraphics[width=\linewidth]{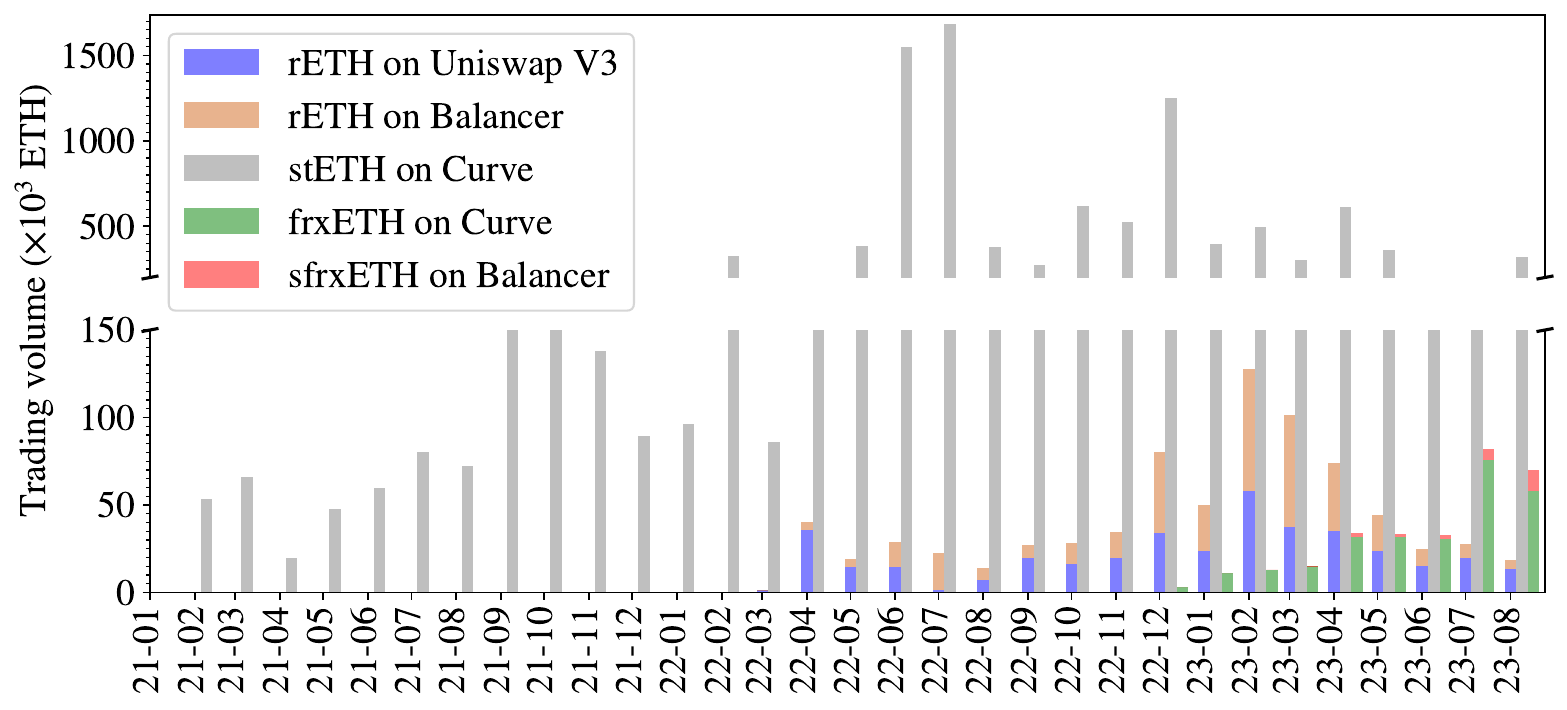}
    \caption{\LSD trading volume over time.}
    \label{fig:lsd_trading_over_time}
\end{minipage}
\end{figure}

\begin{figure}[t]
\centering
\begin{minipage}{.5\textwidth}
   \includegraphics[width=\linewidth]{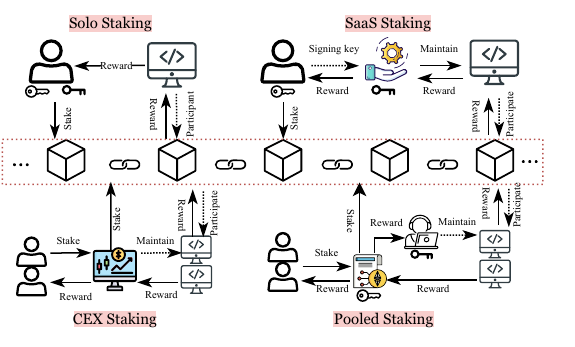}
    \caption{Illustration of staking options on Ethereum.}
    \label{fig:staking_options}
\end{minipage}
\end{figure}

\smallskip
Table~\ref{tab:rocket_arbitrages} summarizes the information of \rETH arbitrages on Unsiwap V3 and Balancer.  Fig.~\ref{fig:staking_options} shows the staking options on Ethereum. Fig.~\ref{fig:rETH_arbitrage_profit_distribution_staking_16eth} depicts the profit distribution for \rETH arbitrages with staking $16$~\ETH. Fig.~\ref{fig:lsd_trading_over_time} shows the distribution of \LSD trading volume on three \DEXs over time.

   \begin{table*}[t]
        \centering
        \resizebox{\textwidth}{!}{ 
        \begin{tabular}{l|c|c|c|c|c|c|c|c|c}
        \toprule
        Contracts	&	\makecell{\# \\ Arbitrageurs}	&	\makecell{\# Arbitrages \\ on Balancer}	&\makecell{\# Arbitrages\\ on UniswapV3}	&\makecell{Arbitrage\\ Profit (\ETH)}	&\makecell{Arbitrage\\ Amount (\ETH)} 	& \makecell{Arbitrage\\ Interval} & \makecell{Code\\ Public} &\makecell{Code\\ Type} &\makecell{Flash\\ Loan}\\
        \hline
        \rowcolor{Gray}
    \href{https://etherscan.io/address/0x1f7e55F2e907dDce8074b916f94F62C7e8A18571}{0x1f7...571} &329 &1406 &497 &1257.8 &30433.5 &2022/11/07-2023/03/07 &\cmark  &Vyper &\cmark \\
\href{https://etherscan.io/address/0x0000007FdaD1dA22cf5A589a513242B50c324071}{0x000...071} &1 &124 &631 &159.8 &14117.9 &2022/12/09-2023/04/16 &\xmark  &- \\
\href{https://etherscan.io/address/0x0000000023191c8382251C0a1AE2f4Db983D414C}{0x000...14C} &1 &311 &62 &181.0 &6771.4 &2022/10/29-2022/12/02 &\xmark  &- \\
\href{https://etherscan.io/address/0x0055Ae46f700BcC53B1b00483d64000d47007200}{0x005...200} &2 &0 &194 &16.1 &2761.5 &2022/10/28-2023/04/17 &\xmark  &- \\
\rowcolor{Gray}
\href{https://etherscan.io/address/0xE46BFe6F559041cc1323dB3503a09c49fb5d8828}{0xE46...828} &24 &80 &28 &79.3 &1807.1 &2023/01/24-2023/04/18 &\cmark  &Vyper &\cmark \\
\href{https://etherscan.io/address/0xa03e47F1B9d7bfb530B5C2AAD073FBC727cBdf61}{0xa03...f61} &1 &0 &95 &13.8 &1519.2 &2023/01/17-2023/02/03 &\xmark  &- \\
\href{https://etherscan.io/address/0x00000000000747D525E898424E8774F7Eb317d00}{0x000...d00} &4 &81 &10 &61.3 &1482.6 &2022/11/01-2023/04/13 &\xmark  &- \\
\href{https://etherscan.io/address/0x7FB3c98821176410d162Df4aC07d7a506CA3Ca52}{0x7FB...a52} &1 &33 &19 &28.8 &1292.8 &2022/10/18-2022/10/29 &\xmark  &- \\
\href{https://etherscan.io/address/0x29492C8Db05DBFfA614222B4Bd9DB646Bb118C72}{0x294...C72} &2 &22 &14 &20.3 &551.5 &2022/12/13-2023/03/12 &\xmark  &- \\
\rowcolor{Gray}
\href{https://etherscan.io/address/0x6fCfE8c6e35fab88e0BecB3427e54c8c9847cdc2}{0x6fC...dc2} &8 &0 &22 &1.3 &483.8 &2023/03/30-2023/04/18 &\cmark  &Solidity &\cmark\\
\href{https://etherscan.io/address/0x6C6B87D44d239B3750bf9BAdce26a9a0A3d2364e}{0x6C6...64e} &1 &0 &20 &0.5 &229.9 &2022/11/03-2022/12/01 &\xmark  &- \\
\href{https://etherscan.io/address/0x1c073d5045b1aBb6924D5f0f8B2F667b1653a4c3}{0x1c0...4c3} &1 &19 &0 &292.8 &4418.8 &2023/04/18-2023/08/07 &\xmark  &- \\
\href{https://etherscan.io/address/0xb7Ed85E4207DDd84d0D9bdD2a4A485DB3A78B74b}{0xb7E...74b} &1 &17 &0 &13.6 &271.9 &2023/02/04-2023/02/20 &\xmark  &- \\
\href{https://etherscan.io/address/0x7f0b35Cf568f914c4880cED7C92c3Ec42418500d}{0x7f0...00d} &1 &0 &17 &1.1 &271.9 &2023/01/09-2023/01/13 &\xmark  &- \\
\href{https://etherscan.io/address/0x7001beAB783Ddd9b68ab073578bDc3A65F7FDd6d}{0x700...d6d} &1 &0 &15 &1.5 &239.9 &2023/02/11-2023/02/15 &\xmark  &- \\
\href{https://etherscan.io/address/0x589d865d1827Eb1dDb7f00049714bf0aB76AB88F}{0x589...88F} &1 &0 &12 &0.2 &304.3 &2022/10/26-2022/11/02 &\xmark  &- \\
\href{https://etherscan.io/address/0xA7b1013E89ed0CC9e96F12BB0aed8D01D458AE3f}{0xA7b...E3f} &1 &6 &5 &5.9 &175.9 &2023/03/15-2023/03/25 &\xmark  &- \\
\href{https://etherscan.io/address/0x16D5a4089CFc8F81E80741aC228245470DD1411c}{0x16D...11c} &1 &0 &10 &0.2 &357.3 &2023/04/16-2023/04/18 &\xmark  &- \\
\href{https://etherscan.io/address/0xA9008da1B57a5C0036DA00bA00009F000054f5d1}{0xA90...5d1} &1 &8 &1 &5.2 &144.0 &2022/11/03-2022/11/06 &\xmark  &- \\
\href{https://etherscan.io/address/0x05A405BE2db49909E2E20B730fd09bC6a55a9d9E}{0x05A...d9E} &1 &0 &8 &0.9 &127.9 &2023/02/20-2023/03/07 &\xmark  &- \\
\href{https://etherscan.io/address/0x91A7cA5c4d7eABC7C8f878Dc49eED3A4Cb28Ea8e}{0x91A...a8e} &1 &0 &7 &0.8 &135.0 &2022/12/04-2023/03/15 &\xmark  &- \\
\href{https://etherscan.io/address/0x86eCD42504FEAD01409901F41cda45a0F8c54e71}{0x86e...e71} &2 &6 &0 &129.9 &2161.5 &2023/04/05-2023/04/10 &\xmark  &- \\
\href{https://etherscan.io/address/0x1A6C2b0D930c8Cb471F6649628bdf6E8c5796d8a}{0x1A6...d8a} &1 &6 &0 &4.8 &96.0 &2023/01/30-2023/02/02&\xmark  &- \\
\rowcolor{Gray}
\href{https://etherscan.io/address/0x321c7DB15c33b93DD907fB2803a23275123D5c7b}{0x321...c7b} &1 &4 &0 &12.5 &343.8 &2022/09/19-2022/09/23 &\cmark  &Solidity &\cmark\\
\href{https://etherscan.io/address/0xF332720d9DaDED8Da18740851dC38B84D0d74742}{0xF33...742} &2 &0 &3 &0.2 &374.8 &2023/05/13-2023/05/20 &\xmark  &- \\
\href{https://etherscan.io/address/0x414783dDd0463F3BD20860B8773E46625ffA7507}{0x414...507} &1 &3 &0 &2.4 &48.0 &2023/02/27-2023/03/05 &\xmark  &- \\
\href{https://etherscan.io/address/0xCCf865Eb42eE041FEA34818c45f2dbe6b7E54923}{0xCCf...923} &1 &3 &0 &23.2 &337.7 &2023/06/11-2023/06/15 &\xmark  &- \\
\href{https://etherscan.io/address/0xc6874B63d09F22b84d835a668720A4c4022e1ee4}{0xc68...ee4} &1 &0 &3 &0.5 &48.0 &2023/02/17-2023/02/18 &\xmark  &- \\
\href{https://etherscan.io/address/0xcb3702bc25B0F284b032e5edf1a1EbeA2FE43255}{0xcb3...255} &2 &0 &2 &0.2 &362.9 &2023/05/12-2023/05/14 &\xmark  &- \\
\href{https://etherscan.io/address/0xEEEeEEEedC84CF6C897c92a02d7E31E6524C935e}{0xEEE...35e} &2 &2 &0 &46.4 &649.7 &2023/07/12-2023/08/01 &\xmark  &- \\
\href{https://etherscan.io/address/0x6C63A4DF7551472145C6bB35BBE91d291129Cd45}{0x6C6...d45} &1 &1 &0 &6.5 &174.9 &2022/09/23-2022/09/23 &\xmark  &- \\
\href{https://etherscan.io/address/0x0c1D5873Fbb668626bf3b4238CECD66747f0Bd05}{0x0c1...d05} &1 &1 &0 &8.3 &119.9 &2023/06/23-2023/06/23 &\xmark  &- \\
\href{https://etherscan.io/address/0x43a3687d47DBD52f76F7cf9795B2a8FCe39BB420}{0x43a...420} &1 &1 &0 &4.5 &124.9 &2022/09/20-2022/09/20 &\xmark  &- \\
\href{https://etherscan.io/address/0xD050E0A4838D74769228B49dFf97241b4Ef3805d}{0xD05...05d} &1 &0 &1 &0.1 &16.0 &2022/12/02-2022/12/02 &\xmark  &- \\
\href{https://etherscan.io/address/0xf35be9a9A0D4B67e776115fA3B6381986379894D}{0xf35...94D} &1 &1 &0 &2.0 &32.0 &2023/04/18-2023/04/18 &\xmark  &-\\
        \bottomrule
        \end{tabular}
        }
        \caption{Contract addresses for \rETH-\ETH arbitrage bots from Sep 30th, 2021, to Aug 10th, 2023.}
        \label{tab:rocket_arbitrages}
    \end{table*}


\end{document}